\documentclass{article}
\usepackage[most]{tcolorbox}
\usepackage{graphicx} 
\usepackage{url}
\usepackage{algorithm}
\usepackage{algpseudocodex}
\usepackage{algorithm,algcompatible,amsmath, relsize}
\usepackage{hyperref}
\usepackage{mathrsfs}
\usepackage[utf8]{inputenc}
\usepackage[T1]{fontenc}
\usepackage{subfig}
\usepackage{multirow}
\usepackage{amsthm,amssymb,inputenc,titling}
\usepackage{orcidlink}
\usepackage{caption}
\usepackage[left=3cm, right=3cm, top=2cm, bottom=2cm]{geometry}
\title{SRAG: RAG with Structured Data Improves Vector Retrieval}
\author{Shalin Shah\\ \small Anvai AI\\ \small shalin@anvai.ai 
        \and 
        Srikanth Ryali\\ \small Anvai AI\\ \small srikanth@anvai.ai
        \and
        Ramasubbu Venkatesh\\ \small Anvai AI\\ \small venky@anvai.ai
        }\date{}
\begin{document}
\maketitle
\begin{abstract}
Retrieval Augmented Generation (RAG) provides the necessary informational grounding to LLMs in the form of chunks retrieved from a vector database or through web search. RAG could also use knowledge graph triples as a means of providing factual information to an LLM. However, the retrieval is only based on representational similarity between a question and the contents. The performance of RAG depends on the numeric vector representations of the query and the chunks. To improve these representations, we propose Structured RAG (SRAG), which adds structured information to a query as well as the chunks in the form of topics, sentiments, query and chunk types (e.g., informational, quantitative), knowledge graph triples and semantic tags. Experiments indicate that this method significantly improves the retrieval process. Using GPT-5 as an LLM-as-a-judge, results show that the method improves the score given to answers in a question answering system by 30\% (p-value = 2e-13) (with tighter bounds). The strongest improvement is in comparative, analytical and predictive questions. The results suggest that our method enables broader, more diverse, and episodic-style retrieval. Tail risk analysis shows that SRAG attains very large gains more often, with losses remaining minor in magnitude.
\end{abstract}
\emph{Keywords}: Retrieval Augmented Generation (RAG), Structured RAG, Structured Metadata, Semantic Tags, Knowledge Graph Triples, Sentiments, Topics, Query Decomposition, Chunk and Query Classes, Financial Question Answering, Representation Learning and Embeddings, SRAG
\section{Introduction}
Retrieval augmented generation (RAG) \cite{lewis2020retrieval} is a widely known method which retrieves information through retrieval from external sources and passes in that information to an LLM \cite{vaswani2017attention} \cite{brown2020language} through prompting. RAG uses vector databases to store the external information, and retrieves the information through vector search. This provides informational grounding, though not epistemic grounding \cite{quattrociocchi2025epistemologicalfaultlineshuman}. We propose a method that uses tagging without requiring changes to the underlying system architecture. The only thing that needs a change is in the way chunks are added to a vector database. We conjecture that SRAG supports episodic-style retrieval and may facilitate latent generalization.\\\\
Our method does not induce learning in the parametric sense, nor does it alter the model’s internal knowledge representations. Instead, we position Structured RAG as supporting in-context generalization by enabling episodic-style retrieval that makes previously encountered information available in context, where it can be flexibly reused. Recent work in \cite{lampinen2025latent} argues that many generalization failures in large language models arise not because the necessary information is absent, but because it is latent-encoded in prior experiences yet inaccessible without reinstating those experiences into context. Episodic retrieval complements parametric learning by converting such latent information into usable in-context evidence, allowing models to apply reasoning procedures they already possess in novel task settings. Our results are consistent with this perspective: by augmenting both queries and chunks with structured metadata, we systematically alter retrieval behavior in a way that promotes broader, more diverse, and semantically aligned retrieval. The resulting improvements on analytical, comparative, and predictive queries suggest that Structured RAG facilitates flexible reuse of information that would otherwise remain inaccessible under similarity-only retrieval.\\\\
Broad and diverse retrieval supports episodic retrieval by increasing the chance that a system reinstates the right past experience, rather than only the most surface-level similar chunk. Generalization failures arise when information learned under one task or format cannot be reused for a different future task because the relevant experience is absent from context. Our method promotes broad and diverse retrieval by augmenting both queries and chunks with structured metadata, such as topics, query types, semantic tags, and knowledge graph triples, which shifts retrieval from purely embedding similarity toward structural, relational, and task-level alignment. This reduces over-filtering by lexical similarity and surfaces experiences that are relevant in how they can be used, not just what they say. By reinstating these latent-bearing experiences into context, the model’s in-context reasoning mechanisms can flexibly recombine prior knowledge, approximating episodic retrieval and mitigating generalization failures that parametric learning alone cannot resolve.\\\\
There have been attempts at adding structured data to standard RAG. The recent work in \cite{yousuf2026utilizing} adds metadata to the chunks and finds similar results, but we add many more types of structured data to the chunks. Furthermore, our claim is stronger through significance testing and tail risk analysis. The work in \cite{pal2025tagging} uses tagging but only at inference time i.e., the tags are passed to an LLM in the prompt, without any need for re-chunking. The methods in \cite{koshorek2025structuredragansweringaggregative} create a structured representation of the corpus and at inference time, it creates structured queries to query the knowledge base. The methods in \cite{jia2025structrag} use a knowledge graph for structured RAG, and they use a knowledge graph for structured retrieval, resulting in a significant infrastructural effort. \cite{shah2024multi} describes a method that uses knowledge graphs as a separate component to retrieve triples through a database and pass them to an LLM through a prompt.\\\\
Our method differs from these, in that, our method does not require architectural changes i.e., the vector database component does not need any modifications. Our method requires re-chunking, but no other infrastructure change is required. Our proposed method uses tagging of both queries and chunks. The chunks are augmented with these tags in the form of semantic metadata. At inference time, the question is also tagged. Rest of the system remains unaltered i.e., the tagged query retrieves the tagged chunks. The tagged chunks are added to the prompt and the LLM synthesizes the answer.\\\\
We tag the queries and chunks with sentiments, topics, semantic tags, knowledge graph triples and the query and chunk classes. The classes are things such as quantitative, comparative, analytical, information lookup, informational etc., which help semantic separation of the chunks. Moreover, we add KG triples to the chunks as well as the query which helps needle in a haystack problems, as well as reasoning problems (e.g., the analytical and comparative query classes) \cite{tan2024struct}.\\\\
Our methods improve comparative, analytical and predictive queries by a large magnitude, indicating that it could be helping LLMs reason better. By tagging the query as well as the chunks, we add metadata (key value pairs as tags) which improves retrieval though better representations. Our method might also improve search and information retrieval based on LLMs \cite{zhu2025large}.\\\\
Here are example queries for which our method is significantly superior (with a GPT-5.2 classification of the \textbf{episodic retrieval impact}):
\begin{itemize}
    \item How does Apple’s AI and on-device intelligence strategy compare to peers such as Microsoft, Google, and Amazon? (\textbf{Impact: High})
	\item How does Apple’s free cash flow conversion compare to large-cap technology peers? (\textbf{Impact: High})
	\item How does Apple’s valuation multiple compare to its historical average? (\textbf{Impact: High})
	\item How does Apple’s exposure to discretionary consumer spending affect revenue growth volatility? (\textbf{Impact: High})
	\item What are the key risks to achieving the upper end of Apple’s FY2025 guidance? (\textbf{Impact: High})
	\item How does Apple’s valuation sensitivity change with USD movements against major currencies? (\textbf{Impact: Medium})
	\item How do productivity and automation improvements affect Apple’s hiring plans and workforce levels? (\textbf{Impact: Medium})
	\item What is the expected impact of seasonality on Apple’s fiscal 3Q FY2025 revenues? (\textbf{Impact: Medium})
	\item What is the expected contribution of major product launches and platform initiatives to FY2026 revenue growth? (\textbf{Impact: Medium})
	\item How does Apple’s free cash flow conversion compare to net income in 1HFY25? (\textbf{Impact: Medium})
	\item What is the expected change in Apple’s working capital in FY2025E? (\textbf{Impact: Low})
	\item What is the expected employee attrition or workforce stability trend for Apple in FY2025E? (\textbf{Impact: Low})
	\item What is Apple’s projected return on invested capital (RoIC) for FY2027E? (\textbf{Impact: Low})
	\item What is the expected operating (EBIT) margin for Apple in FY2027E? (\textbf{Impact: Low})
	\item What is the expected EPS for Apple in FY2026E? (\textbf{Impact: Low})
\end{itemize}
\begin{table}[t]
\caption{LLM-as-a-Judge GPT-5 Scores (Scores in [0, 100]) (Welch's independent t-test with unequal variances)}
\label{results1}
\begin{center}
\begin{tabular}{p{4cm}p{2cm}p{3cm}p{2cm}}
\hline
\bf Query Class &\bf Plain RAG &\bf Structured RAG &\bf p-value\\
\hline
All Queries &72.36 &\bf 94.35 &2e-13\\
\hline
Predictive &64.46 &\bf 95.61 &9e-5\\
\hline
Information Lookup &\bf 98.37 &97.43 &0.24\\
\hline
Analytical &65.1 &\bf 93.8 &2e-5\\
\hline
Comparative &55.9 &\bf 94.1 &0.0002\\
\hline
Informational &83.3 &\bf 93.5 &0.03\\
\hline
Quantitative &80.5 &\bf 94.1 &0.04\\
\hline
\end{tabular}
\end{center}
\end{table}

\begin{figure}[ht]
\label{barchart_1}
\centering
\includegraphics[scale=0.4]{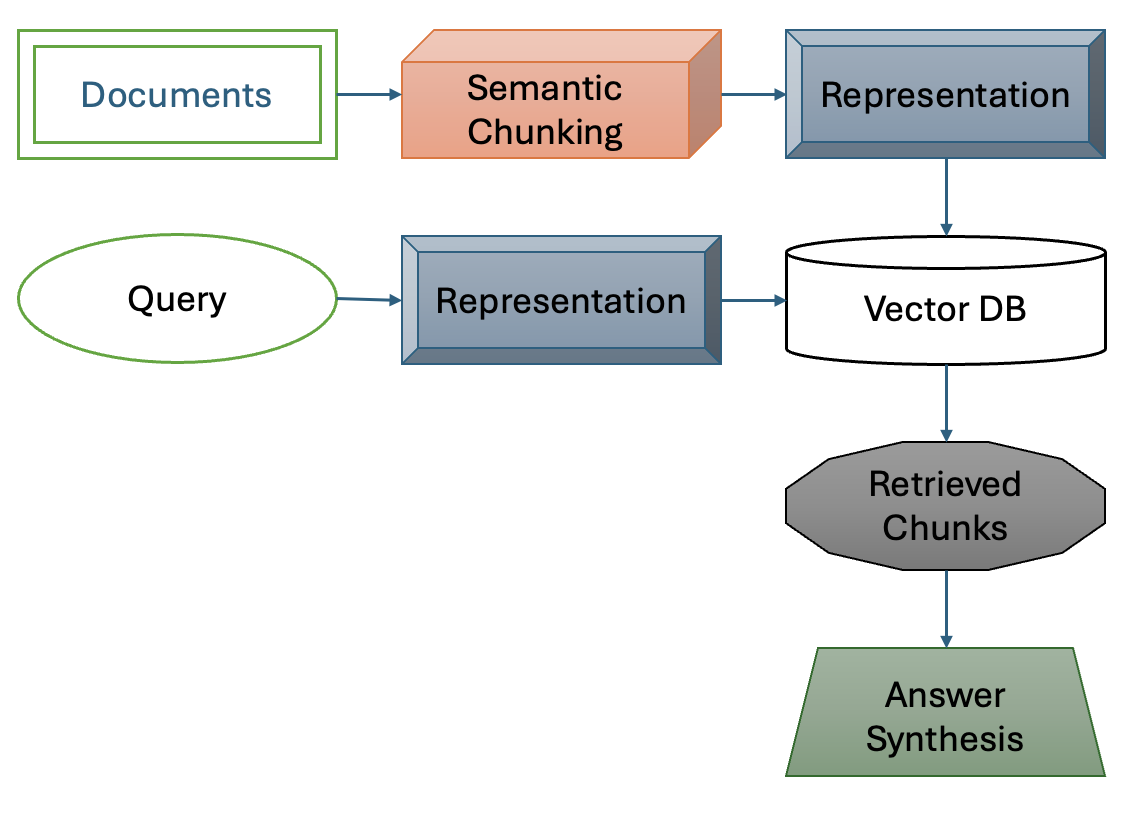}
\caption{Plain RAG (PRAG)}
\end{figure}

\begin{figure}[ht]
\label{barchart_2}
\centering
\includegraphics[scale=0.35]{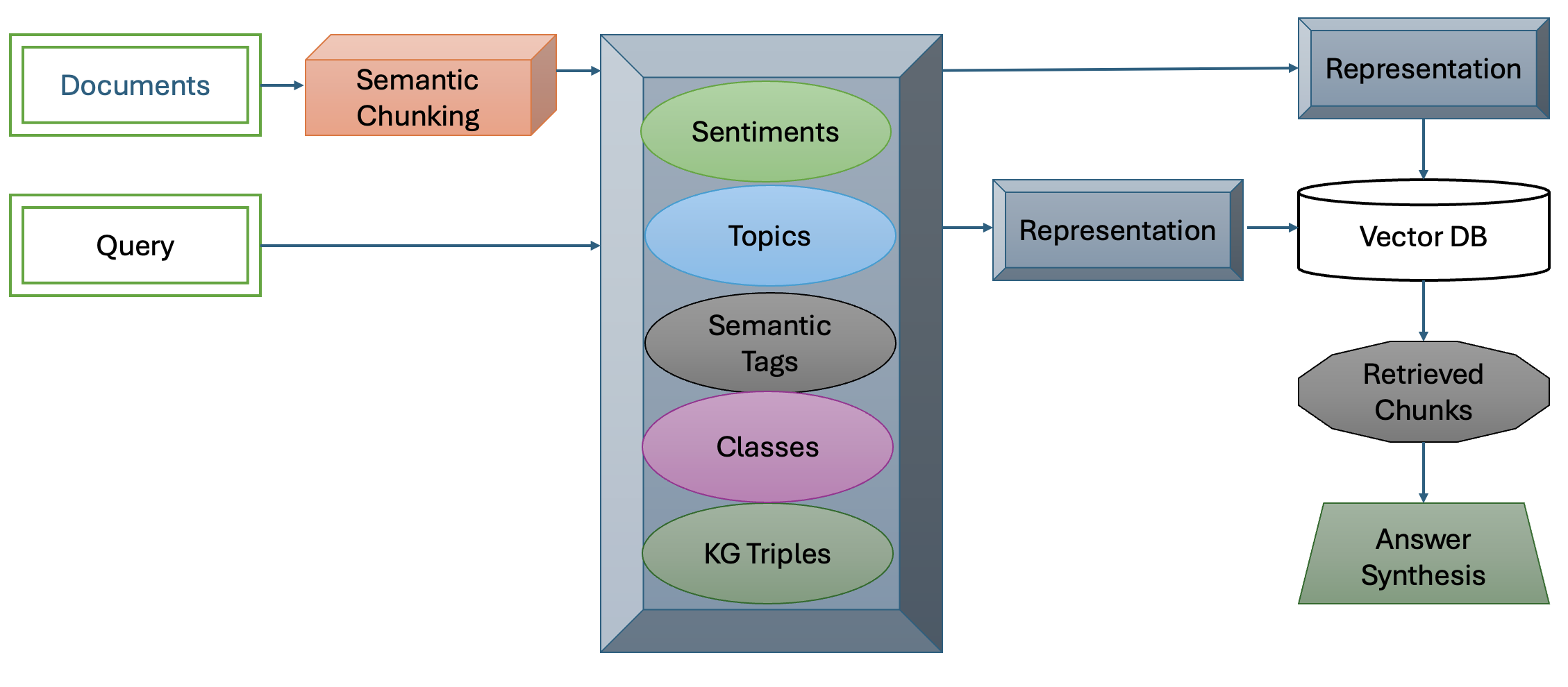}
\caption{Structured RAG (SRAG)}
\end{figure}

\section{Key Contributions}
There have been attempts at using structured information in retrieval augmented generation methods, but all of them (to our knowledge) require infrastructural changes (e.g., a graph database). Our contributions are as follows:
\begin{itemize}
    \item We propose using structured data, appended to all chunks
    \item Our structured data includes semantic rags, sentiments, query and chunk classes, topics and knowledge graph triples
    \item Our method outperforms vanilla RAG on all query classes except the information\_lookup class, for which the performance of both methods is similar
    \item Our conjecture is that our method, to some extent, supports better generalization through retrieval through broad, diverse and episodic retrieval
    \item Our empirical observations show that our method significantly improves informational grounding through a very simple to implement method
\end{itemize}
\section{Results}
Table 1 and figures 3 and 4 show the results of our methods. Our methods improve the LLM-as-a-Judge (GPT-5) scores by 30\%. Considering classes of queries and chunks, our methods improve the scores of analytical, comparative and predictive queries very strongly, indicating an improvement in reasoning capabilities. The information lookup class has a neutral effect, but this could be due to the fact that information lookup queries are less in the distribution (figure 5).\\\\
The results indicate that the retrieval of chunks in our methods promotes diversity, and can retrieve information that is broadly related to the query. By adding the sentiment, and tags, our method might help episodic retrieval. Diversity is important here, while still maintaining strong semantic and factual relevance. The complete thought process of an LLM is guided by diversity and broadness, without reducing the performance on needle in a haystack questions, such as information lookup. Moreover, the performance on informational and quantitative queries is also superior to plain RAG.\\\\
The ablation over the number of retrieved chunks (figure 6) highlights an important property of Structured RAG. We observe that performance gains are more pronounced at smaller values of k, indicating that the method improves early retrieval precision rather than relying on larger contexts to compensate for retrieval noise. When only a few chunks are retrieved, vanilla RAG is more likely to miss critical context, whereas Structured RAG surfaces semantically and structurally aligned chunks earlier due to improved representation shaping. As k increases, both methods benefit from higher recall, and the relative advantage of Structured RAG naturally diminishes as the language model can compensate for imperfect retrieval during generation. These results suggest that Structured RAG reduces dependence on large retrieval budgets by improving the quality and relevance of top-ranked results.
\begin{figure}[ht]
\label{barchart_4}
\centering
\includegraphics[scale=0.5]{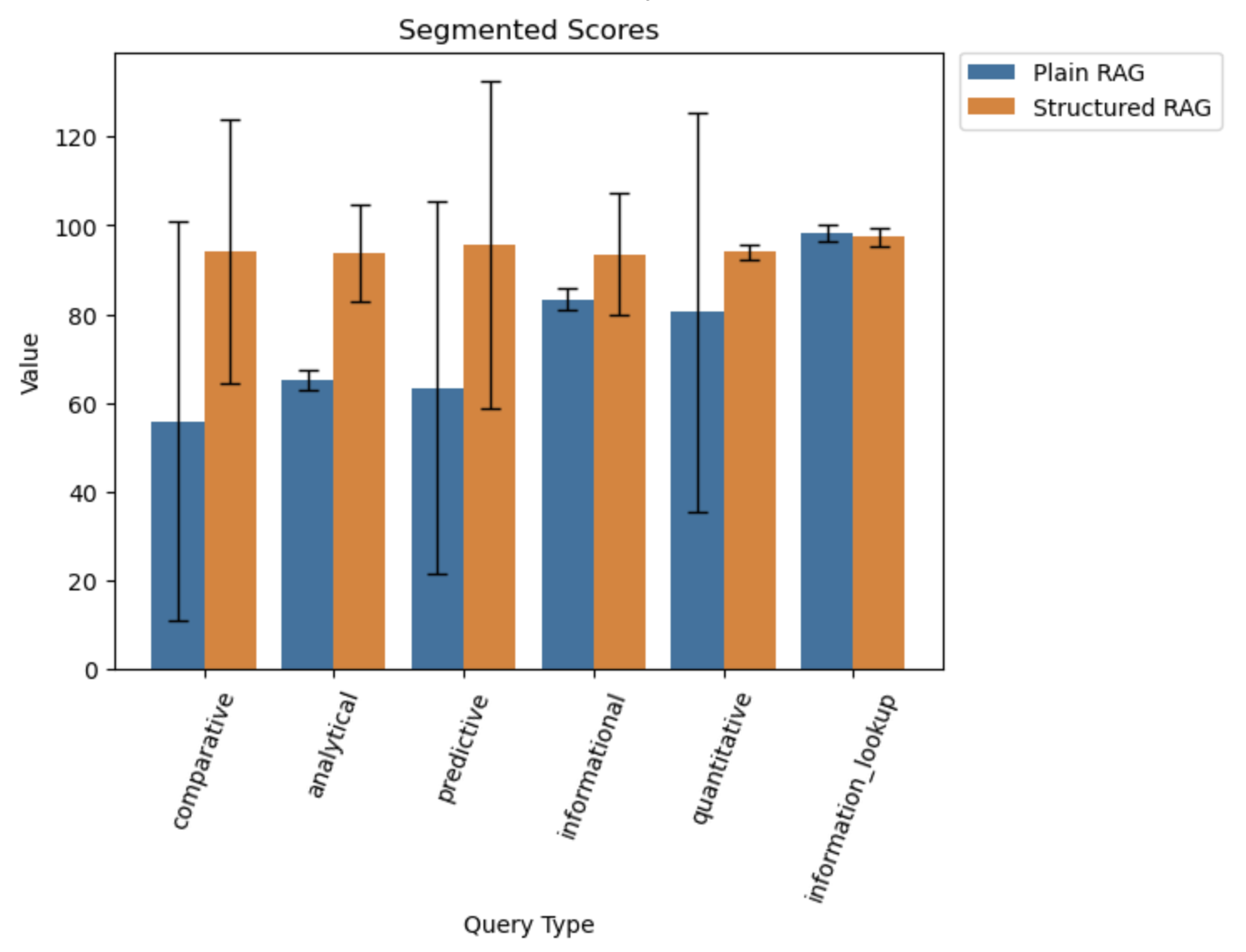}
\caption{Scores across different classes of queries (Scores in [0, 100])}
\end{figure}
\begin{figure}[ht]
\label{barchart_5}
\centering
\includegraphics[scale=0.35]{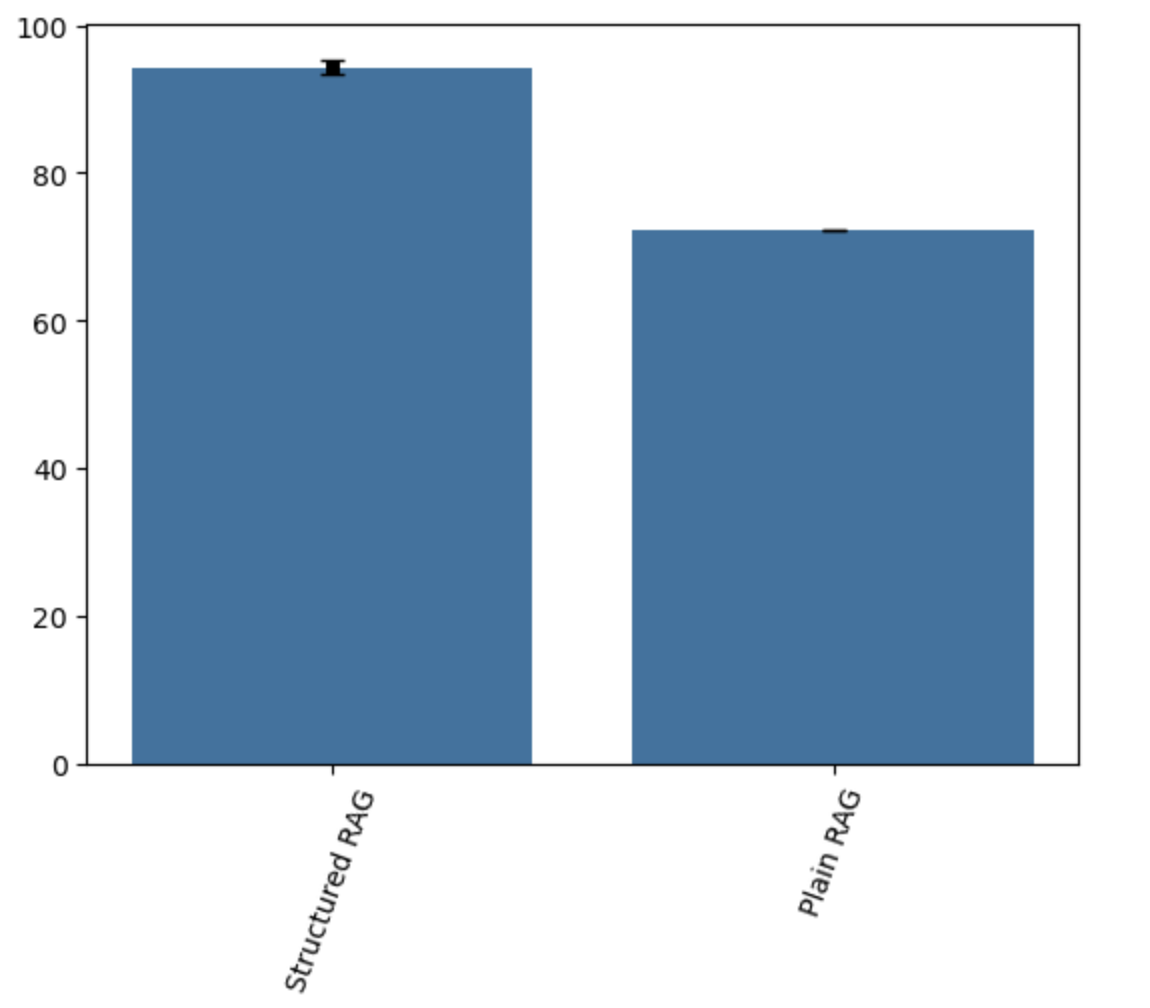}
\caption{Total Scores (Scores in [0, 100])}
\end{figure}
\begin{figure}[ht]
\label{barchart_6}
\centering
\includegraphics[scale=0.35]{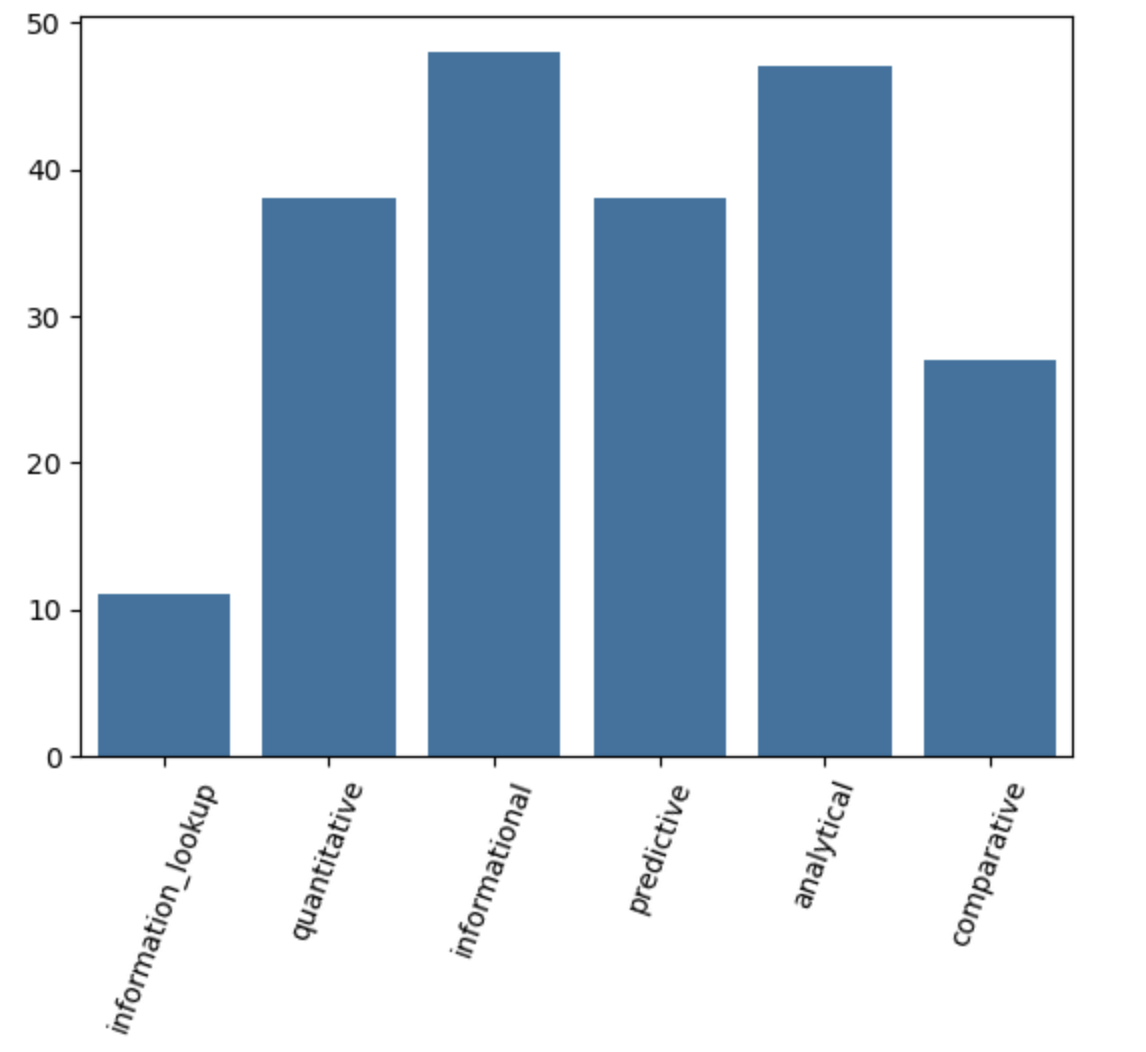}
\caption{Query class distribution}
\end{figure}
\section{Ablation Study}
The marginal ablation results show that removing individual metadata components does not lead to statistically significant performance changes when considered in isolation, even though the full Structured RAG system yields large and statistically significant gains across most query classes. This pattern might be typical in ablation studies of modern ML systems, where improvements arise from interactions among multiple correlated components rather than from any single feature acting independently. Within this context, there is still a clear ordering in the magnitude of the effects; semantic tags, topics, and chunk type exhibit the largest negative score changes when ablated, while KG triples and sentiment have near zero impact. This aligns with the overall results, where Structured RAG substantially improves analytical, comparative, predictive, informational, and quantitative queries; tasks that depend heavily on correct semantic framing and intent alignment; while showing no significant gain on information lookup queries.\\\\
The lack of statistical significance in the marginal ablations should therefore not be interpreted as evidence that these components are unimportant. Instead, it reflects several well understood properties of ablation analysis; partial redundancy among metadata signals, compensatory effects when one signal is removed, and evaluation noise dominating small deltas. Such behavior might be common and expected in ablation studies, particularly when the system is robust and its gains are distributed rather than brittle. Additionally, some metadata, most notably sentiment, is intrinsically less relevant for factual financial queries, which further limits its marginal impact when ablated. Taken together, these results support the interpretation that Structured RAG’s improvements are emergent and compositional, driven primarily by the joint use of semantic tags, topics, and chunk types, with individual ablations appearing statistically insignificant precisely because no single component is solely responsible for the observed gains.\\\\
Ideally, a principled attribution of each metadata component’s contribution would require evaluating the full power set of feature subsets and computing Shapley values, which measure average marginal contributions across all possible coalitions. However, this approach was computationally and financially infeasible in our setting. Exhaustively evaluating all $2^k$ subsets would require re-running retrieval, generation, and LLM-based evaluation for each configuration, resulting in prohibitive compute cost and latency. As a result, we restrict our analysis to marginal ablations, which provide directional insight while acknowledging that they cannot fully capture interaction effects among metadata components.
\begin{figure}%
    \centering
    \subfloat[\centering k=3]{{\includegraphics[width=4.5cm]{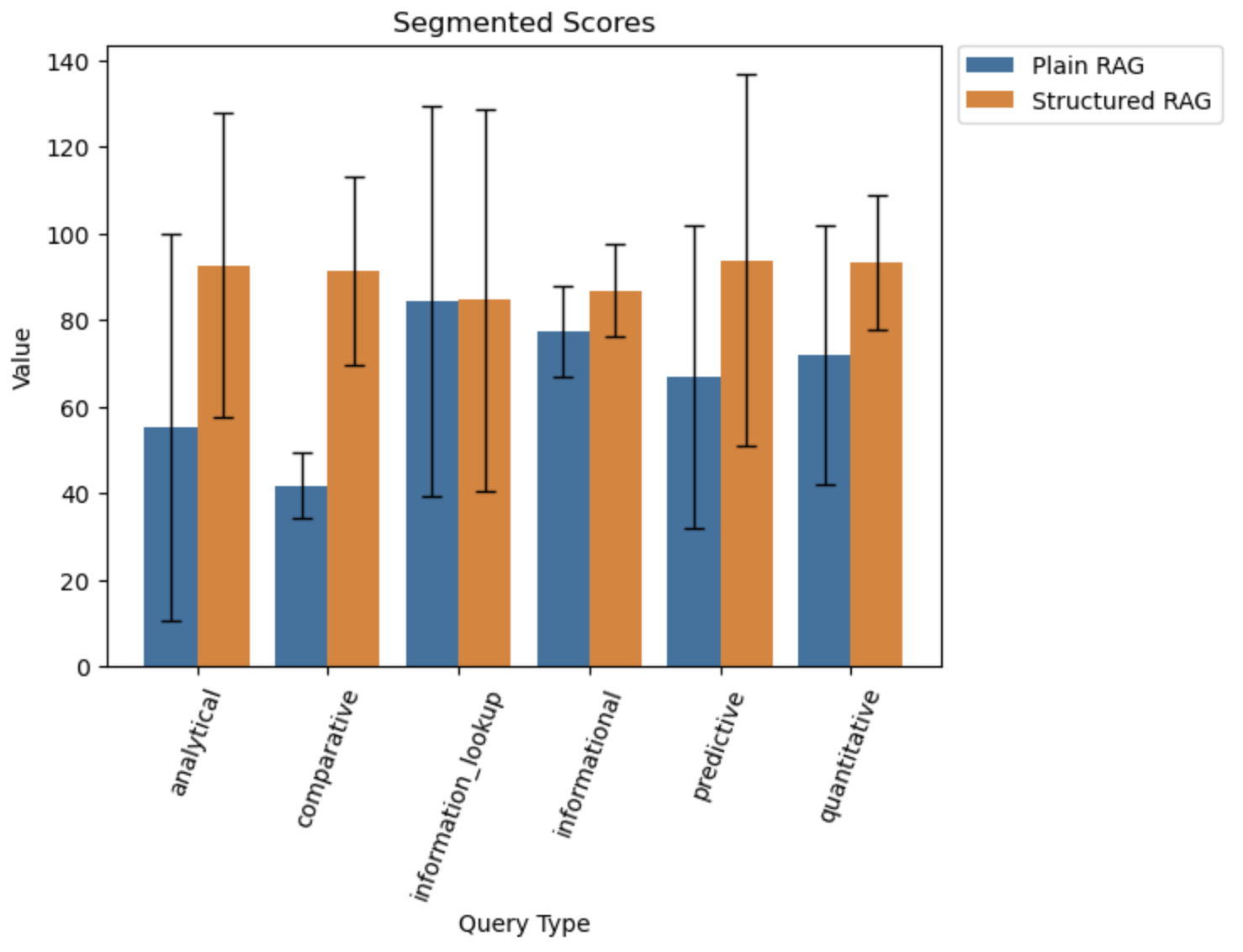}}}%
    \qquad
    \subfloat[\centering k=5]{{\includegraphics[width=4.5cm]{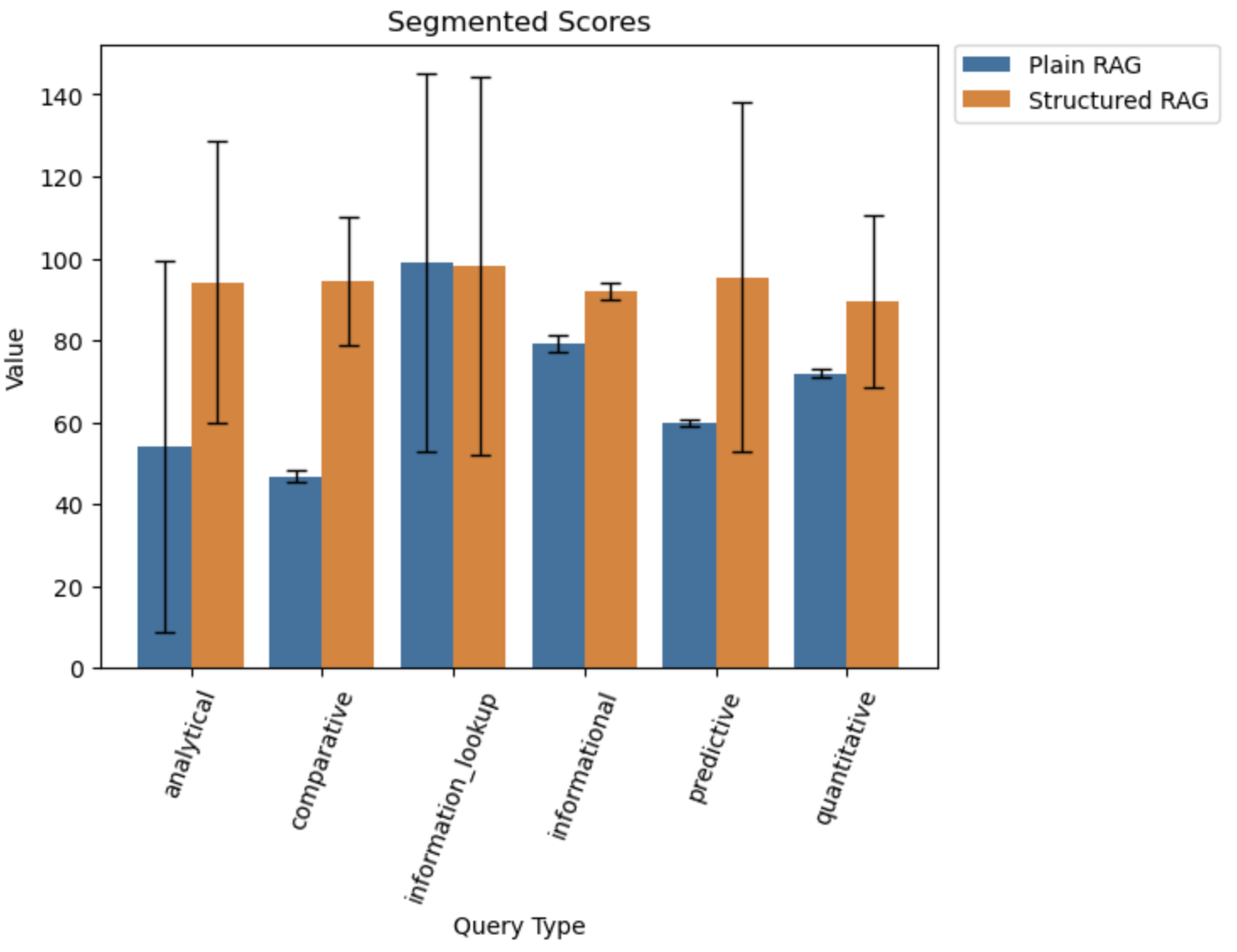} }}
    \qquad
    \subfloat[\centering k=10]{{\includegraphics[width=4.5cm]{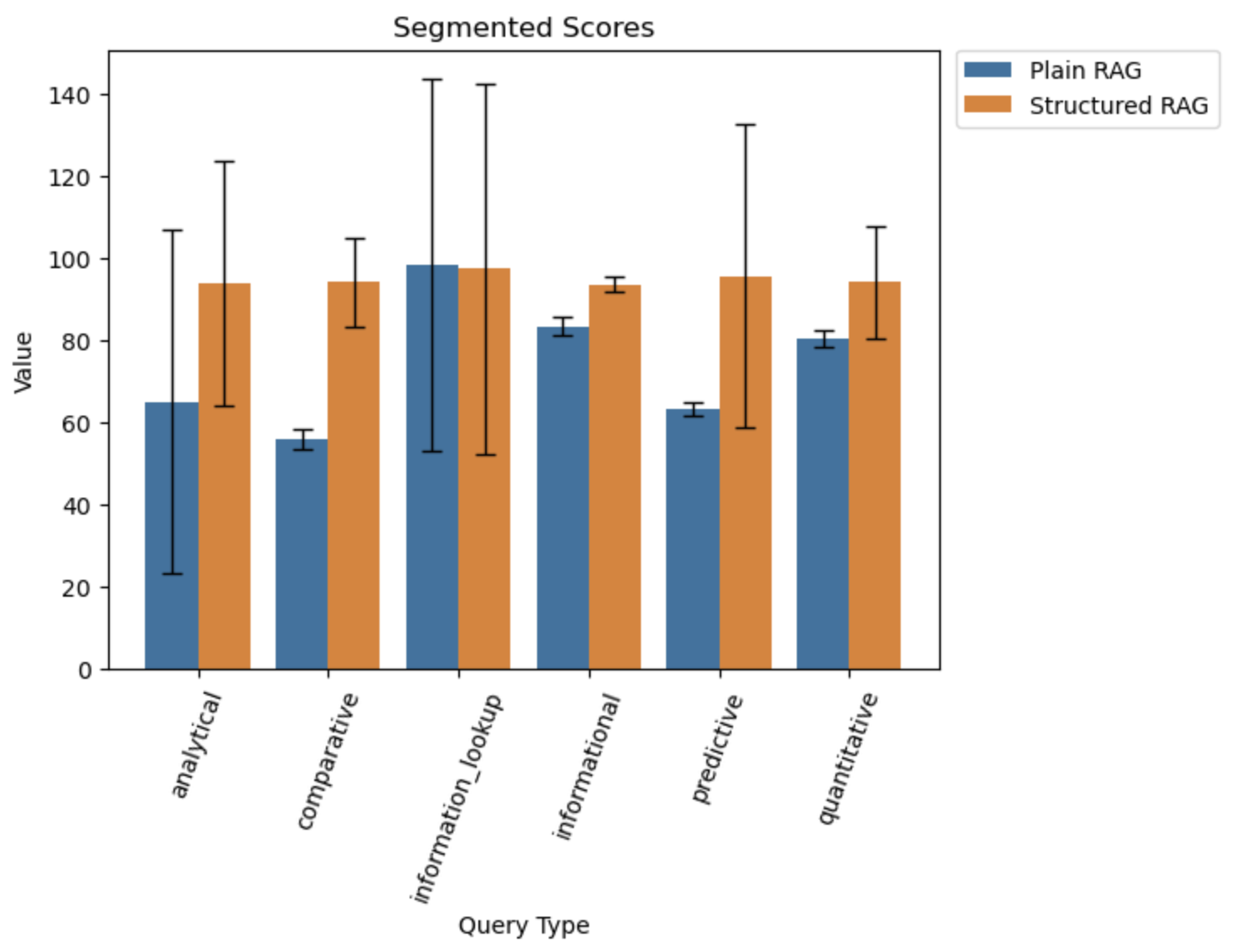} }}%
    \caption{Ablation by k (number of retrieved chunks)}%
    \label{fig:example}%
\end{figure}
\begin{table}[t]
\caption{Ablation Study (Marginal Contributions) (Scores in [0, 100])}
\label{results2}
\begin{center}
\begin{tabular}{p{4cm}p{3cm}p{2cm}}
\hline
\bf Metadata Ablated &\bf Change in Score &\bf p-value\\
\hline
No Semantic Tags &-1.1 &0.15\\
\hline
No Topics &-0.53 &0.55\\
\hline
No Chunk Type &-0.3 &0.7\\
\hline
No KG Triples &0.01 &0.99\\
\hline
No Sentiment &0.63 &0.37\\
\hline
\end{tabular}
\end{center}
\end{table}
\begin{table}[t]
\caption{Tail Risk Analysis (SRAG - PRAG) (Scores in [0, 100])}
\label{results15}
\begin{center}
\begin{tabular}{p{5cm}p{2cm}}
\hline
\bf Name &\bf Value\\
\hline
5th Percentile Difference &\bf -2.9\\
\hline
10th Percentile Difference &-2.5\\
\hline
90th Percentile Difference &\bf 88.1\\
\hline
Median Difference &0.0\\
\hline
Max Loss &\bf -7.9\\
\hline
P(SRAG Better) &0.441\\
\hline
P(SRAG Worse) &0.407\\
\hline
Average Harm When Worse &\bf -1.95\\
\hline
Average Gain When Better &\bf 51.76\\
\hline
\end{tabular}
\end{center}
\end{table}
\begin{figure}[ht]
\label{barchart_3}
\centering
\includegraphics[scale=0.4]{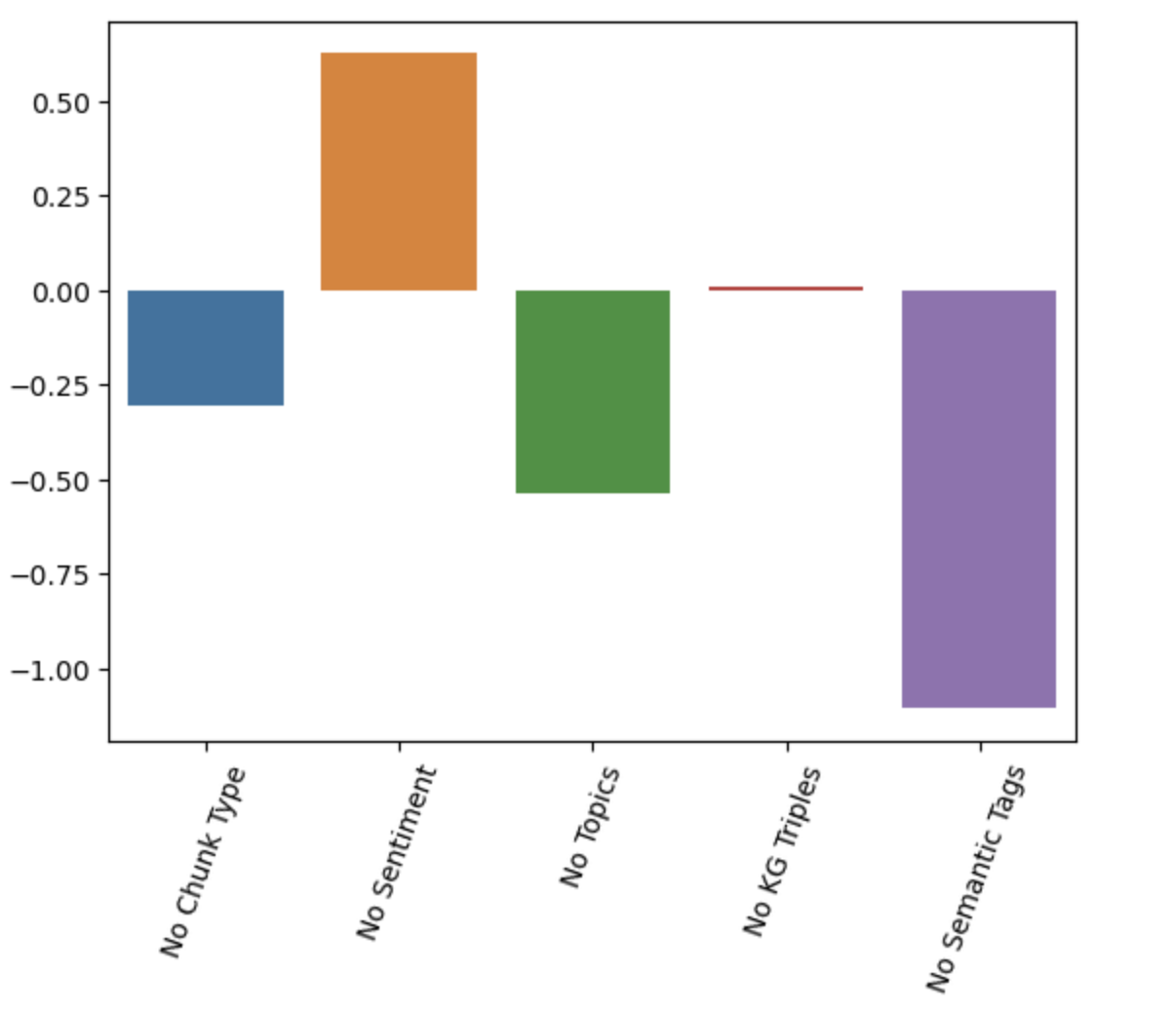}
\caption{Ablation Study}
\end{figure}
\section{Conclusion}
We presented a simple and practical extension to standard Retrieval Augmented Generation by augmenting both queries and chunks with structured metadata prior to indexing in a vector database. Unlike prior structured RAG approaches, our method does not require architectural or infrastructural changes such as graph databases or hybrid retrievers. Instead, it relies on re-chunking and tagging to reshape the underlying vector representations used for retrieval. This design choice makes the approach easy to integrate into existing RAG pipelines while directly targeting one of the core limitations of vanilla RAG; its reliance on surface-level representational similarity between queries and chunks.\\\\
Empirically, our results show a substantial improvement of approximately 30\% (p-value 2e-13) over plain RAG using an LLM-as-a-judge evaluation, with particularly strong gains for analytical, comparative, and predictive queries. These classes of queries typically require aggregation, comparison, or forward-looking reasoning across multiple pieces of information, where similarity only retrieval might fail. Also, SRAG has tighter confidence intervals as compared to plain RAG. The observed improvements suggest that augmenting representations with structured signals such as topics, query and chunk types, semantic tags, and knowledge graph triples promotes broader, more diverse, and task-aligned retrieval. Interpreted through the lens of recent work on latent learning \cite{lampinen2025latent}, our findings are consistent with the view that improved retrieval can support in-context generalization by reinstating relevant prior experiences into context, enabling language models to flexibly reuse reasoning capabilities they already possess.\\\\
Finally, our ablation study provides insight into the role of different metadata components while highlighting the limitations of single feature ablations in systems with interacting signals. The results suggest that performance gains are emergent and compositional, arising from the joint use of multiple forms of structured metadata rather than from any single dominant feature. While a full power set analysis with a Shapley value attribution would be required for a complete decomposition of feature contributions, such an analysis was computationally infeasible in our setting. Future work could explore more principled attribution methods, extend evaluation to additional domains, and compare this lightweight approach with a full knowledge-graph–augmented retrieval system to better understand the tradeoffs between infrastructure complexity and retrieval effectiveness.
\bibliography{main}
\nocite{*}
\bibliographystyle{unsrt}
\appendix
\section{Appendix}
\begin{tcolorbox}
\textbf{Illustration 1: Metadata included in a chunk}\\\\
--- METADATA ---\\\\
Type: financial\_table\\\\
Sentiment: positive\\\\
Topics: Revenue Growth, Earnings Estimates, Market Performance, Investment Recommendation, Capital Allocation, Cash Flow Generation\\\\
KG Triples:
Apple -> reported -> resilient quarter with mid-single-digit y/y revenue growth;
Apple -> maintained -> FY2025 revenue growth outlook in low-to-mid single digits;
Apple -> raised -> Fair Value to \$220;
Apple -> net income -> \$97 bn declined modestly y/y due to margin normalization;
Apple -> operating margin -> stable at ~30\%\\\\
Tags:
{‘FY2025 EPS estimate change’: ‘0–2\%’,
‘Fair Value’: ‘\$220’,
‘CMP’: ‘\$195’,
‘52-week range’: ‘\$199–\$164’,
‘Market Cap’: ‘\$3,000 bn’,
‘ADTV-3M’: ‘\$12.5 bn’,
‘EPS 2025E’: ‘7.10’,
‘EPS growth 2025E’: ‘8.5\%’,
‘P/E 2025E’: ‘27.5’,
‘Sales 2025E’: ‘\$410 bn’,
‘Net profits 2025E’: ‘\$101 bn’,
‘Div. yield 2025E’: ‘0.6\%’}
\end{tcolorbox}
\end{document}